\documentstyle[aps,epsf,amssymb,twocolumn,floats]{revtex}

\begin{document}
\draft
\title{ Multiple-path interferometer with a single quantum obstacle }
\author{H. Schomerus$^{1}$, Y. Noat$^2$, J. Dalibard$^3$, and C. W. J.
Beenakker$^4$}
\address{$^1$Max-Planck-Institut f\"ur Physik komplexer Systeme, N\"othnitzer
Str. 38,
01187 Dresden, Germany\\
$^2$Kamerlingh Onnes Laboratory, Universiteit Leiden, P.O. Box 9504, 2300 RA
Leiden, The Netherlands\\
$^3$Laboratoire Kastler Brossel,
D\'epartement de Physique de l'Ecole Normale Sup\'erieure,
24 rue Lhomond, 75231 Paris, France
\\
$^4$Instituut-Lorentz, Universiteit Leiden, P.O. Box 9506, 2300 RA
Leiden, The Netherlands
}
\date{\today}
\twocolumn[
\widetext
\begin{@twocolumnfalse}
\maketitle
\begin{abstract}
We consider the scattering of particles by 
an obstacle which tunnels coherently between two
positions.
We show that the obstacle mimics two classical scatterers
at fixed positions when the kinetic energy $\varepsilon$ of the incident
particles is smaller than the tunnel splitting $\Delta$:
If the obstacles are arranged in parallel, 
one observes an interference pattern as in the conventional
double-slit experiment.
If they are arranged in series, the observations
conform with a Fabry-Perot interferometer.
At larger $\varepsilon$ inelastic processes result in 
more complex interference phenomena.
Interference
disappears when $\varepsilon \gg \Delta$, but can be recovered 
if only the elastic scattering channel is detected.
We discuss the realization of a quantum obstacle in mesoscopic systems.
\end{abstract}
\vspace{0.5cm}
\narrowtext
\end{@twocolumnfalse}
]
\narrowtext

In  the familiar double-slit
experiment a beam of particles is sent through two slits in 
a plate
and the transmitted intensity is observed on a screen.
One finds an interference pattern, thus demonstrating
the coherent superposition of the two possible scattering paths.
The same result is obtained for the reflection from obstacles,
e.\,g., the bars of a reflection grating.

\begin{figure}
\epsfxsize7cm 
\centerline{\epsfbox{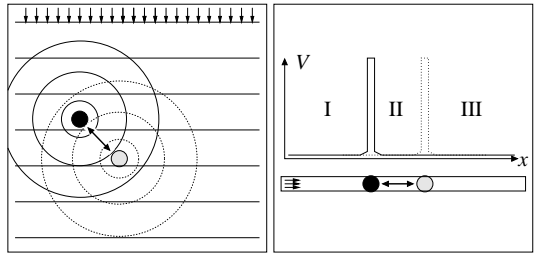}}
\medskip
\caption{Scattering of a particle
by a delocalized obstacle which tunnels between two
positions (solid and shaded dot).
Left panel: double slit, the obstacle provides two scattering paths
in parallel. The lines indicate fronts of the wave function.
Right panel: Fabry-Perot interferometer, the obstacle's positions
are arranged in series. The solid and dashed curves indicate the
scattering potentials.
The problem is solved by matching the waves in the regions I, II, and III.
} 
\label{fig1}
\end{figure}

In these experiments the slits or bars only play a passive role.
In this paper we consider scattering by a
single obstacle, which, however, is by itself a quantum object with states
that correspond to different locations (see Fig.\ \ref{fig1}).
Allowing for superpositions
of these states, we have an obstacle that is delocalized in space.
Which reflection or transmission pattern is then observed on the screen?
What we will demonstrate here is that the
quantum obstacle may act as a collection of classical scatterers, in that
one observes the corresponding interference pattern on the screen.
The condition for interference is that the tunnel splitting $\Delta$ 
is of order or larger than
the kinetic energy $\varepsilon$ of the incident particles.
This suggests the interpretation that
the obstacle has to tunnel quickly enough in order to give
the particles a choice of the path.
For $\Delta\ll \varepsilon$ the
interference pattern from the obstacle
disappears, but can be recovered
if only the elastic scattering channel is detected.
(It is also recovered
by coincidence
detection of the final state of the obstacle 
in its delocalized eigenbasis.)

The double slits or bars are two scattering elements put in parallel.
We contrast this with the one-dimensional problem of two
barriers arranged in series---a Fabry-Perot interferometer. 
The quantum analogue 
is an obstacle which tunnels between two locations along the
propagation direction of the incident particles.
We solve exactly the one-dimensional model with a repulsive contact interaction
(delta-function potential $U\delta(x)$) and find transmission resonances
for $\Delta\gtrsim \varepsilon$, hence again that the quantum obstacle
acts as two classical obstacles if it tunnels quickly enough.
In the limit of large interaction strength $U$ and sufficiently large
separation between the positions of the obstacle, the 
transmission amplitude becomes identical to
the transmission amplitude of the Fabry-Perot interferometer.
In general, the transmission probability remains finite even if the
barrier strength $U$ is infinite. 
At smaller tunnel splitting one finds a rich behavior of the
transmission probability due to multiple
inelastic scattering.
These results are further illuminated by 
the delay time, a measure of the interaction time \cite{delaytimes}.

Let us first consider the `parallel' quantum scatterer
in three dimensions, which hops between two
positions ${\bf R}_1={\bf r}_0/2$ (internal state $|1\rangle$),
${\bf R}_2=-{\bf r}_0/2$ (internal state $|2\rangle$),
with separation
${\bf r}_0$.
The eigenstates of the scatterer are the symmetric
and antisymmetric combinations
$|s\rangle=2^{-1/2}(|1\rangle+ |2\rangle)$
and $|a\rangle=2^{-1/2}(|1\rangle- |2\rangle)$,
respectively. The corresponding eigenenergies
${\mp}\Delta/2$ differ by the tunnel splitting $\Delta>0$.
(As usual we assume that the ground state is the symmetric state
$|s\rangle$.)
This gives
rise to tunneling of the scatterer
between the two positions at a frequency $\nu=\Delta/h$.
The incident particle with coordinate ${\bf r}$
interacts with the scatterer at position ${\bf R}_k$ 
through a potential $V_k({\bf r})=V({\bf r}-{\bf R}_k)$ which
depends only on its relative position to the scatterer.
The Hamiltonian describing the total system
composed of the particle and the scatterer can then be written as
\begin{eqnarray}
H&=&
\frac{{\hat{\bf p}}^2}{2m}+\hat W
- \frac{\Delta}{2}\left[
|1\rangle\langle 2|+|2\rangle\langle 1| \right],
\nonumber\\
\hat W&=&V_1({\bf r}) |1\rangle\langle 1| +V_2({\bf r})
|2\rangle\langle 2|.
\label{eq:ham}
\end{eqnarray}
Here $\hat {\bf p}$ is the momentum operator
of the particle of mass $m$. The plane-wave eigenstates $|{\bf k}\rangle$
are denoted by their wave vector ${\bf k}={\bf p}/\hbar$.

We assume that the scatterer is initially prepared in its ground
state $|s\rangle$
(preparation in its excited state $|a\rangle$ is equivalent
to the case $\Delta<0$;
superpositions result in nonstationary behavior).
In the limit of a weak interaction we
can apply the
Born approximation  
and obtain the probability of scattering 
from the initial
state $|{\bf k}_{\rm i}\rangle$
into the final state
$|{\bf k}_{\rm f}\rangle$ by summing the probabilities for
each final state of the scatterer:
\begin{eqnarray}
P_{{\bf k}_{\rm i} \rightarrow {\bf k}_{\rm f} }&=&\frac{2\pi}{\hbar}\Big\{
\left|\langle {\bf k}_{\rm f}; s|\hat W|s;{\bf
k}_{\rm i}\rangle\right|^2\delta \left(\varepsilon_{\rm i}-\varepsilon_{\rm f}
\right)
\nonumber\\ &&{}
+\left|\langle {\bf k}_{\rm f}; a|\hat W|s;{\bf
k}_{\rm i}\rangle\right|^2 \delta \left(\varepsilon_{\rm i}-\varepsilon_{\rm
f}-\Delta
\right)\Big\}.
\end{eqnarray}
For of a short-ranged potential of the form
$V({\bf r})=U\delta \left({\bf r}\right)$ the probability reads
\begin{eqnarray}
P_{{\bf k}_{\rm i} \rightarrow {\bf k}_{\rm f}}&=&\frac{2\pi}{\hbar} U^2
[\cos^2\left(\Delta {\bf k}\cdot{\bf r_0}/2\right)
\delta\left(\varepsilon_{\rm i}-\varepsilon_{\rm f} \right)
\nonumber\\ &&{}
+ \sin^2\left(\Delta {\bf k}\cdot{\bf
r_0}/2\right)\delta\left(\varepsilon_{\rm i}-\varepsilon_{\rm f}-\Delta\right)]
,
\label{eq:born}
\end{eqnarray}
where $\Delta {\bf k}={\bf k}_{\rm i}-{\bf k}_{\rm f}$.
Interference with full contrast
is observed when the energy $\varepsilon_i$ of the incoming
particle is smaller than $\Delta$, because then 
the
argument of the second delta function is always negative
(inelastic processes are forbidden).
In this situation the quantum
scatterer acts as two classical scatterers of fixed positions ${\bf R}_1$
and ${\bf R}_2$,
since $P_{{\bf k}_{\rm i} \rightarrow
{\bf k}_{\rm f}}\sim \cos^2\left(\Delta {\bf k}\cdot{\bf r}_0\right/2)$. On
the other hand, interference is lost if the
kinetic energy of the
incident particles $\varepsilon_{\rm i}\gg \Delta$
\cite{ctba}.
The interference pattern can be recovered
if one only detects the elastic scattering channel
(by 
means of energy-resolved detection at energy $\varepsilon$).

Now we turn to the `serial' quantum barrier, which hops in
the propagation direction of the scattered particle. 
In the case of one-dimensional scattering 
(plane-parallel barriers, or confined propagation)
and for the delta-function
potential
$V_1(x)=U\delta(x-L/2)$, $V_2(x)=U\delta(x+L/2)$,
this scattering
problem 
can be solved exactly.
[The problem is defined by the Hamiltonian given in Eq.\ (\ref{eq:ham})
with these potentials and
$\hat {\bf p}^2$ replaced by
$\hat p_x^2=-\hbar^2\partial^2/\partial x^2$.]
In order to simplify the notation we use units
$\hbar^2/2m\equiv 1$, such that the kinetic energy $\varepsilon=k^2$.

Before we present the solution,
let us briefly recall the results for the conventional
case of immobile barriers.
For a single immobile barrier of strength $U$
the transmission and reflection amplitudes at
wave number $k$ are given by 
\begin{equation}
t=\frac{1}{1+iU/2k} \quad\mbox{and}\quad
r=\frac{iU/2k}{1+iU/2k},
\label{eq:deltabarrier}
\end{equation}
respectively,
such that the reflection probability approaches unity when $U\gg |k|$.
When two such immobile barriers are placed in series with separation $L$ 
they form a Fabry-Perot interferometer, with
transmission amplitude
\begin{equation}
t=\frac{k^2}{k^2+ikU+U^2[\exp(2ikL)-1]/4}.
\label{eq:tdimer}
\end{equation}
For a large finesse of the interferometer ($U/k \gtrsim 1$) 
one finds the well-known transmission resonances close to integer values
of $kL/\pi$.

The quantum scatterer is delocalized, giving rise to a number of
additional resonance and interference effects.
In order to explore these effects we
solve the stationary scattering problem for
electrons with momentum $k>0$ incident from the left,
while the scatterer is
prepared in the eigenstate $|s\rangle$, hence giving the total energy
$E=\varepsilon-\Delta/2$.
Under conservation of this energy, the electrons can be 
reflected or transmitted either elastically or inelastically,
where in the latter case the outgoing electrons have momentum
$\pm q$, with $q=\sqrt{\varepsilon-\Delta}$, and the scatterer is
excited into the
state $|a\rangle$.

The scattering probabilities and phase shifts 
can be determined via wave matching of the wavefunctions
\begin{equation}
\phi_\alpha=
(a_{\alpha}e^{ikx}+ b_{\alpha}e^{-ikx})|s\rangle 
+(c_{\alpha}e^{iqx}+d_{\alpha}e^{-iqx})|a\rangle
\end{equation}
at the boundaries of the three regions $\alpha=\rm I$ for $x<-L/2$,
$\alpha=\rm II$ for $-L/2<x<L/2$, and $\alpha=\rm III$ for $x>L/2$
(see Fig.\ \ref{fig1}).
The resulting linear system of equations is then solved
for $b_{\rm I}$, $d_{\rm I}$,
$a_{\rm III}$, and $c_{\rm III}$ as a linear function of $a_{\rm I}$
(which we set to unity),
under the conditions $c_{\rm I}=b_{\rm III}=d_{\rm III}=0$,
because no electrons are coming in 
with momentum $q$, or from the right.
(For the case $\varepsilon<\Delta$ we use the convention ${\rm Im}\,q>0$,
so that the wavefunctions with coefficient $d_{\rm I}$ and
$c_{\rm III}$ decay exponentially with the distance to the scatterer.)
The coefficients (which are lengthy algebraic expressions 
and hence not written down here)
determine the  elastic and inelastic transmission
and reflection probabilities by  $T_{\rm el}=|a_{\rm III}|^2$,
$T_{\rm inel}=({\rm Re}\, q/k)|c_{\rm III}|^2$,
$R_{\rm el}=|b_{\rm I}|^2$, and $R_{\rm inel}=({\rm Re}\, q/k)|d_{\rm I}|^2$.
(The inelastic scattering probability $T_{\rm inel}+
R_{\rm inel}$ vanishes for $\varepsilon<\Delta$.)
The coefficients also deliver
the scattering phase shifts
$\phi_{\rm R,T;el,inel}$, with $\phi_{\rm T;el}={\rm arg}\,a_{\rm III}$,
etc, and the delay times
$\tau=\hbar {\rm d}\phi/{\rm d}\varepsilon$.

Let us now discuss some special cases.
In the limit $\Delta\to 0$ of slow tunneling we find 
\begin{eqnarray}
&&R=\frac{U^2}{4k^2+U^2},\quad R_{\rm el}=\cos^2(kL)R, \quad
R_{\rm inel}=\sin^2(kL)R
,
\nonumber\\
&&T=T_{\rm el}=\frac{4k^2}{4k^2+U^2},\quad T_{\rm inel}=0
.
\label{eq:delta0}
\end{eqnarray}
Qualitatively, the parameter dependence of Eq.\ (\ref{eq:delta0})
corresponds to the result in Born approximation, Eq.\ (\ref{eq:born}),
with parallel vectors ${\bf k}_f\parallel{\bf k}_i\parallel{\bf r}_0$.
Moreover,
the total transmission and reflection probabilities are the same as
for the conventional problem of a single delta function of strength $U$
[see Eq.\ (\ref{eq:deltabarrier})].
However, the scatterer has a finite probability $R_{\rm inel}$ to change its
internal state.
(This can be probed in the elastic channel if
$\Delta$ is greater than the energy
resolution of the detector, but still much smaller than 
$\varepsilon$.)

In the limit $U\gg k$ of strong scattering, according to Eq.\ (\ref{eq:delta0})
the transmittance vanishes for $\Delta=0$.
In striking contrast, a finite transmission probability results for
$\Delta\neq 0$ even in the limiting case  $U\to\infty$,
which can be interpreted as systematic avoidance of the
particle and the scatterer.
For $\Delta<\varepsilon$
we find in this limit the coefficients
\begin{eqnarray}
a_{\rm III}&=&2ik\exp(iqL)g^{-1}\,{\rm Im}\,f ,\nonumber \\
c_{\rm III}&=&\exp[i(k-q)L/2]a_{\rm III} ,\nonumber\\
b_{\rm I}&=&2q g^{-1}[i\,{\rm Im}\,f-k(\cos kL+\cos qL)],\nonumber\\
d_{\rm I}&=&k\exp[-i(k+q)L/2]g^{-1} \nonumber\\
&&\times [(\exp[2iqL]-1)k+(\exp[2ikL]-1)q],
\label{eq:vinf}
\end{eqnarray}
where $f=q\exp(ikL)-k\exp(iqL)$, $g=(k+q)^2-f^2$.
The proportionality between the transmission 
coefficients $a_{\rm III}$  and $c_{\rm III}$
in  Eq.\ (\ref{eq:vinf}) entails for the elastic
and inelastic transmission delay times
the relation
\begin{equation}
\tau_{\rm  T;inel}=\tau_{\rm  T;el}-\frac{L}{4}
\left(\frac{1}{q}-\frac{1}{k}\right)
\frac{2m}{\hbar}
,
\end{equation}
where we momentarily reintroduced the units $\hbar$, $m$.

\begin{figure}
\epsfxsize7cm 
\centerline{\epsfbox{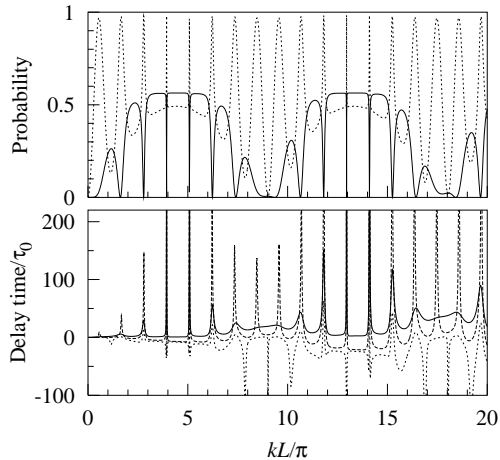}}
\caption{Upper panel:
Probabilities of transmission $T_{\rm el}+T_{\rm inel}$
(solid curve) and inelastic 
scattering $T_{\rm inel}+R_{\rm inel}$
(dashed curve) for the one-dimensional model with $U=\infty$ 
as a function of $kL/\pi$, for $\Delta/\varepsilon=0.4$. Lower panel:
delay times 
$\tau_{\rm T;el}$ (solid curve), $\tau_{\rm R;el}$
(long dashes, peaks pointing upwards), and $\tau_{\rm R;inel}$
(short dashes, peaks pointing downwards), in units of
$\tau_0=\hbar/\varepsilon$.}
\label{fig2}
\end{figure}

The corresponding probabilities of transmission and inelastic-scattering,
as well as the delay times 
$\tau_{\rm T;el}$, $\tau_{\rm R;el}$, and $\tau_{\rm R;inel}$,
are plotted in Fig.\ \ref{fig2} as function
of $kL/\pi$ for fixed $\Delta/\varepsilon=0.4$.
We find a regular sequence of transmission zeros,
accompanied by long delay times for the various scattering processes.
The peaks of $\tau_{\rm R, el}$ point upwards, the peaks
of $\tau_{\rm R, inel}$ point downwards.
The transmission probability is modulated by a
function $F(kL/\pi)$ [related to $f$ in Eq.\ (\ref{eq:vinf})]
with period $p=2/(1-\sqrt{1-\Delta/\varepsilon})$
and maxima at $kL/\pi=(n+1/2)p$. In the limit 
$\varepsilon\gg \Delta$ (where $p\simeq
4\varepsilon/\Delta$) the maxima occur 
when the time of flight $L/v$ of the particle between
the two positions of the obstacle is
an odd multiple
of the tunneling time $h/2\Delta$ 
between these positions (the minima occur at even multiples).
Close to the minima
the peaks in $\tau_{\rm R, el}$ and $\tau_{\rm R, inel}$
alternate; close to the maxima they coincide.

At $\Delta=\varepsilon$ the momentum $q$ vanishes, and the inelastic
scattering rate
(which is proportional to ${\rm Re}\, q$)
drops to zero.
The transmission probability becomes
\begin{equation}
T=T_{\rm el}=\frac{(kL-\sin kL)^2}{(1+\cos kL)^2+(kL-\sin kL)^2}.
\end{equation}
In the limit $kL\gg 1$ the transmission probability $T=1$.
This is a remarkable observation:
The scatterer becomes totally transparent although $U\to\infty$.

Another remarkable case of total transmission is found
for $\Delta>\varepsilon$, where all electrons are scattered elastically.
The transmission probability is now
\begin{eqnarray}
&&T=T_{\rm el}=4k^2(k \sinh L|q|-|q|\sin kL)^2 e^{-2L|q|}
\nonumber\\
&&\times|(k+i|q|)^2 -
e^{ikL}(k e^{-L(|q|+ik)} -i|q| )^2|^{-2}
.
\end{eqnarray}
The delay times $\tau_{\rm T}$ and $\tau_{\rm R}$ are equal (this is a
joint consequence of
the unitarity of the scattering matrix 
and the reflection symmetry of the potential).
At large length $L\gg 1/|q|$ the transmission amplitude $t$
becomes exactly identical to the transmission amplitude of
the Fabry-Perot interferometer,
Eq.\ (\ref{eq:tdimer}), with $U$ replaced by $2|q|$.
Hence, although we started out with an infinite scattering strength
$U$, the finesse of the quantum version of the
Fabry-Perot interferometer is finite.

\begin{figure}
\epsfxsize7cm 
\centerline{\epsfbox{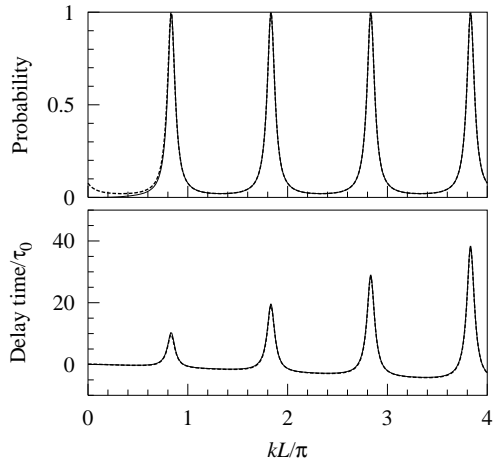}}
\caption{
Transmission probability $T$ (upper panel) 
and delay time $\tau$ in units of $\tau_0=\hbar/\varepsilon$
(lower panel), 
as a function of $kL/\pi$.
The solid curve is the result for the 
quantum obstacle
with scattering strength $U=\infty$ and 
$\Delta/\varepsilon=4$, the dashed curve
(nearly indistinguishable from the solid curve)
the result for
the conventional Fabry-Perot interferometer
(two fixed classical barriers)
from Eq.\ (\ref{eq:tdimer})
with $U/k=2\sqrt{3}$.
}
\label{fig3}
\end{figure}

The transmission probability and the delay time
is plotted for $\Delta/\varepsilon=4$
as a function of $kL/\pi$ in Fig.\ \ref{fig3} (solid curves).
The dashed curves show these quantities for two fixed classical
barriers with scattering strength $U/k=2\sqrt{3}$.
The comparison again demonstrates that the quantum obstacle
behaves as two fixed classical scatterers when the 
tunnel splitting exceeds the kinetic energy.

Let us briefly discuss some implications of our results for mesoscopic
systems.
A possible manifestation of the delocalized scatterer
is an interstitial defect, like a light atom,
which hops between two energetically equivalent positions.
When the thermal excitation energy is of the order of 
the energy barrier that the particle has to overcome, the defect jumps
incoherently from one position to another.
Since the
potential in the system is changed after each jump, the
conductance exhibits random temporal fluctuations 
(telegraphic noise) between two
values $G_1$ and $G_2$ 
\cite{tfnoise1,tfnoise2,tfnoise3}.
For a long measurement time one measures the time average $\langle G
\rangle=(G_1+G_2)/2$.

For lower temperature one enters the coherent regime
in which the defect acts as a quantum obstacle. If $\Delta$ is sufficiently
large  the defect is in its ground state, as we have assumed in our analysis.
Up to now, however, we have neglected many-body effects. 
The most straight-forward modification is to account for
the Pauli blocking of states below the Fermi energy $E_{\rm F}$:
Inelastic scattering is forbidden when the excitation energy
$\varepsilon-E_{\rm F}<\Delta$. 
(The typical excitation energy is given by the potential drop
$eV$ or by the thermal excitation energy, whatever the
larger.)
More intricate many-body effects arise, for example, from
sequential scattering of several particles
by the same quantum scatterer.
If these can be neglected and inelastic processes are forbidden by 
Pauli blocking, we can use the Landauer formula $G=(e^2/h)T$ to relate 
the conductance to the transmission probability calculated above.

Can one also realize the serial quantum obstacle in one
dimension (the Fabry-Perot interferometer)? An experimentally
controllable set-up
could consist of a single-channel wire
placed adjacent to a double-quantum-dot
device, which is tuned in resonance in the Coulomb-blockade regime.
(For some experiments on double dots see Ref.\ \cite{dd1,dd2}.)
The quantum obstacle is the electron which occupies the two
degenerate levels on the dots, and interacts with the
electrons in the wire by Coulomb repulsion. However, once
again for a quantitative theory one cannot neglect many-body effects.
In order to circumvent these complications to some extent, one might think of
injecting ``hot electrons'' from one end of the wire, by shooting them over an
additional potential barrier. In this way the excitation energies can
be restricted to a small energy interval which is well separated from the
Fermi energy.

In summary, we have investigated scattering
by a quantum obstacle which is delocalized in space,
and found close analogies to the double slit and
the Fabry-Perot interferometer.
This owes to the
capability of the scatterer to act as a collection
of classical scatterers for fast coherent tunneling
(which can be interpreted as unsuccessful resolution 
of the scatterer position by the particle). A striking 
feature of the quantum obstacle is that an infinitely
high potential barrier can become transparent
(which can be interpreted as
successful avoidance of the
particle and the scatterer).
Additional regimes with a
rich phenomenology have been identified, depending
on the kinetic energy of the incoming particle and the tunnel
frequency of the scatterer.

Recent experiments on mesoscopic structures have probed the scattering by
tunable \cite {dd1,dd2}
or spontaneously formed \cite{bj1,bj2}
two-level systems consisting of mobile entities.
These are classical obstacles at
high temperatures, but could turn into quantum obstacles as the temperature
is lowered. It is challenging to find
signatures for having entered this new regime.
We propose the appearance of interference effects like the ones 
discussed in this paper as such a signature.
We have concentrated
on a single-particle scenario which neglects many-body effects
of the scattered particles, most importantly 
sequential scattering of several particles
by the same quantum scatterer.
In view of the mesoscopic applications,
our results
could serve as the building block for a more quantitative theory
that includes these effects.
This is a promising subject for further investigations.

We thank S. Bahn, 
H. Bouchiat, A. Georges, Y. Imry, M. Kociak, and B. Reulet 
for discussions.
This work was
supported by the Dutch Science Foundation NWO/FOM
and the European Commission.

\end{document}